\documentclass[aps,pre,twocolumn,showpacs,superscriptaddress,groupedaddress]{revtex4}  
\usepackage{amsfonts}
\usepackage{amsmath}
\usepackage{amssymb}
\usepackage{version}
\usepackage{graphicx}%
\includeversion{New_connection}
\excludeversion{Old_connection}

\begin{document}
\title[Verlet Revision]{Application of the G-JF Discrete-Time Thermostat for Fast and Accurate Molecular Simulations}
\author{Niels Gr{\o}nbech-Jensen}
\affiliation{Department of Mechanical and Aerospace Engineering, University of California, Davis, CA 95616.}
\affiliation{Department of Chemical Engineering and Materials Science, University of California, Davis, CA 95616.}
\affiliation{Computational Research Division, Lawrence Berkeley National Laboratory, Berkeley, CA 94720.}
\author{Natha Robert Hayre}
\affiliation{Department of Physics, University of California, Davis, CA 95616.}
\author{Oded Farago}
\affiliation{Department of Mechanical and Aerospace Engineering, University of California, Davis, CA 95616.}
\affiliation{Department of Biomedical Engineering, Ben Gurion University of the Negev, Be'er Sheva, 84105 Israel.}
\affiliation{Ilse Katz Institute for Nanoscale Science and Technology, Ben Gurion University of the Negev, Be'er Sheva, 84105 Israel.}

\keywords{Molecular Dynamics, Verlet Algorithm, Simulated Langevin Dynamics, Stochastic Differential Equations}
\pacs{02.70.Ns, 02.70.-c, 05.10.-a, 02.60.Cb} \today

\begin{abstract}
A new Langevin-Verlet thermostat that preserves the fluctuation-dissipation relationship
for discrete time steps, is applied to molecular modeling and tested against several popular
suites (AMBER, GROMACS, LAMMPS) using a small molecule as an example that can be easily simulated
by all three packages.
Contrary to existing methods, the new thermostat exhibits no detectable changes in the sampling statistics
as the time step is varied in the entire numerical stability range. The simple form of the method, which we express in the three common forms
(Velocity-Explicit, St{\"o}rmer-Verlet, and Leap-Frog), allows
for easy implementation within existing molecular simulation packages to achieve faster and more accurate results with no
cost in either computing time or programming complexity.
\end{abstract}
\maketitle

\section{Introduction}
Recently, a new stochastic thermostat \cite{G-JF}, based on an exact implementation of the
fluctuation-dissipation relationship in discrete time, was presented as integrated
into the well-known and widely used Verlet formalism. It was analytically demonstrated
that the method provides exact thermodynamic response for both
flat and harmonic potentials for any time step within the Verlet stability criteria. This unique thermodynamic feature of the approach,
as implemented into the Verlet framework, makes it attractive for a wide range of applications, where it is desired to
execute efficient and accurate simulations. Here, we demonstrate that the G-JF method of Ref.~\cite{G-JF} is not limited to simple linear cases, but also extends its usefulness to complex nonlinear systems.

Langevin dynamic simulations constitute an appealing approach for simulations of physical systems in contact with
a thermodynamic heat bath. A very popular class of such systems is molecular dynamics (MD) \cite{Frenkel_1996}
of classical particle ensembles. The method is based on numerical integration of the Langevin equation
\begin{eqnarray}
m\dot{v} & = & f(r,t)-\alpha\dot{r}+\beta(t) \, , \label{eq:Eq_Intro_2a}
\end{eqnarray}
where $r$ is the coordinate, $v=\dot{r}$ is its velocity, $m$ is its mass, $f$ is a net deterministic force acting
on $r$. The friction constant $\alpha>0$ and the noise $\beta$ are connected by the fluctuation-dissipation
relationship \cite{Parisi_1988}
\begin{eqnarray}
\langle\beta(t)\rangle & = & 0\label{eq:Eq_Intro_3a}\\
\langle\beta(t)\beta(t^\prime)\rangle & = & 2\alpha k_BT\delta(t-t^\prime)\label{eq:Eq_Intro_4a}
\end{eqnarray}
with $k_B$ and $T$ being Boltzmann's constant and the thermodynamic temperature, respectively.
Given the applicability of this equation to a wide spectrum of physical systems with phenomenological
dissipation and thermal noise, there have been decades of development in numerical methods for
solving such equations. Specifically, the perfection of the most important thermodynamic properties
of discrete-time numerical simulations have been of particular interest. We here point to the vast literature through
Refs.~\cite{G-JF,LM,Brunger_1984,Loncharich_2004,vdSpoel_2005,Goga_2012,Schneider_1978,vGunsteren_1988,V-EC}
and references therein.
The most commonly sought after properties in stochastic simulations have been {\it i}) diffusion of a particle in a flat
potential, {\it ii}) transport on a linear ramp potential (which can be mapped directly onto the diffusive behavior in a flat potential), and {\it iii}) Boltzmann sampling in harmonic potentials \cite{notice_on_methods}. Since the G-JF method \cite{G-JF} exhibits all these 
features thermodynamically correct in discrete time, we here wish to provide additional expressions for practical implementation
of the method. We also provide a simple, yet representative, example of how the method performs for a nonlinear and complex system
in comparison to several widely used contemporary molecular dynamics simulation suites.

\section{The Three Verlet Expressions}
Verlet integrators are usually expressed in one of three forms \cite{Allen}: {\rm A}) velocity-explicit Verlet
(VE), which advances the trajectory one time step
based on the coordinate and its conjugate variable (here the velocity), {\rm B}) St{\"o}rmer-Verlet (SV), which uses
coordinates at two consecutive time steps to advance time, and {\rm C}) leap-frog (LF), which advances the
trajectory based on the coordinate and its conjugate variable, the latter being represented at half-integer
time steps relative to the coordinate.
The three typical Verlet formulations produce {\it identical} trajectories, and, thus,
are different expressions of the exact same method. Consequently,
applications of the Verlet method are commonly expressed in any of the available forms.
Since all three identical methods
are frequently used in the literature, we start here with
the recently published G-JF thermostat that was derived
in the natural VE form, and re-express the algorithm in
the two other popular forms. The previous harmonic
oscillator analysis \cite{G-JF} of the method applies to all three
variants since they produce identical results. Specifically,
the three following formulations will result in the correct
fluctuation-dissipation relationship and thus, the correct
thermodynamic response in linear systems.

\subsection{Velocity Explicit G-JF}
We take our starting point with the VE G-JF expressions that were derived and analyzed in \cite{G-JF}.
Denoting discrete time variables by the integer time step superscript, such that, e.g., $r^n=r(t_n)$, the
algorithm for advancing $r^n$ and $v^n$ one time step of $dt$ reads
\begin{eqnarray}
r^{n+1} & = & r^{n}+b\, dt \, v^{n} + \frac{b\, dt^2}{2m}f^{n} + \frac{b\, dt}{2m}\beta^{n+1}\label{eq:Eq_VE_1a}\\
v^{n+1} & = & av^{n}+\frac{dt}{2m}(af^{n}+f^{n+1})+ \frac{b}{m}\beta^{n+1} \, , \label{eq:Eq_VE_2a}\end{eqnarray}
where
\begin{eqnarray}
a & = & \frac{1-\frac{\alpha dt}{2m}}{1+\frac{\alpha dt}{2m}} \; \; , \; \; 
b \; = \; \frac{1}{1+\frac{\alpha dt}{2m}}\, , \label{eq:Eq_VE_4a}
\end{eqnarray}
and where
\begin{eqnarray}
\beta^{n+1} & = & \int_{t_n}^{t_{n+1}}\beta(t^\prime) \, dt^\prime  \label{eq:Eq_VE_5a}
\end{eqnarray}
is a standard Gaussian random number that satisfies
\begin{eqnarray}
\langle\beta^n\rangle & = & 0 \; \; , \; \; 
\langle\beta^n\beta^l\rangle \; = \; 2\alpha k_BTdt\delta_{n,l} \, .\label{eq:Eq_VE_6a}
\end{eqnarray}
Setting the initial conditions $(r^0,v^0)$, Eqs.~(\ref{eq:Eq_VE_1a})-(\ref{eq:Eq_VE_6a}) can be directly used to generate the trajectory $(r^n,v^n)$
from which the dynamical and statistical information can be derived.

\subsection{St{\"o}rmer-Verlet G-JF}
We here start by rewriting Eqs.~(\ref{eq:Eq_VE_1a}) and (\ref{eq:Eq_VE_2a}) for $n\rightarrow n-1$:
\begin{eqnarray}
r^{n} & = & r^{n-1}+b\, dt \, v^{n-1} + \frac{b\, dt^2}{2m}f^{n-1} + \frac{b\, dt}{2m}\beta^{n}\label{eq:Eq_VE_1b}\\
v^{n} & = & av^{n-1}+\frac{dt}{2m}(af^{n-1}+f^{n})+ \frac{b}{m}\beta^{n} \, . \label{eq:Eq_VE_2b}
\end{eqnarray}
As outlined in Ref.~\cite{G-JF}, Eq.~(\ref{eq:Eq_VE_2b}) can be inserted into Eq.~(\ref{eq:Eq_VE_1a}) in order to
replace $v^n$, whereafter Eq.~(\ref{eq:Eq_VE_1b}) is used to replace the resulting $v^{n-1}$.
This yields
\begin{eqnarray}
r^{n+1} & = & 2br^n-ar^{n-1}+\frac{b\,dt^2}{m}f^n+\frac{b\;dt}{2m}(\beta^n+\beta^{n+1}) \, , \nonumber \\ \label{eq:Eq_SV_1a}
\end{eqnarray}
which is the SV formulation of the G-JF method. Unlike the VE expressions, the SV equation
does not contain direct information about the velocity and is therefore not directly applicable
for natural initial conditions $(r^0,v^0)$. The
self-consistent approach for starting this procedure from $(r^0,v^0)$ is to apply Eq.~(\ref{eq:Eq_VE_1b}) for $n=0$,
\begin{eqnarray}
r^1 & = & r^0+b\, dt \, v^0 + \frac{b\, dt^2}{2m}f^0 + \frac{b\, dt}{2m}\beta^{1} \, , \label{eq:Eq_SV_2a}
\end{eqnarray}
then apply Eq.~(\ref{eq:Eq_SV_1a}) for all subsequent time steps $n>0$. In order to calculate both the complete
dynamical trajectory and important thermodynamic quantities, one needs the velocity $v^n$ explicitly
expressed as well. Consistent with the method, we replace $v^{n-1}$ in Eq.~(\ref{eq:Eq_VE_2b}) by 
inserting Eq.~(\ref{eq:Eq_VE_1b}), such that
\begin{eqnarray}
v^n & = & \frac{a}{b}\frac{r^n-r^{n-1}}{dt}+\frac{dt}{2m}f^{n}+\frac{1}{2m}\beta^{n} \nonumber \\
& = & \frac{r^{n+1}-(b-a)r^n-ar^{n-1}}{2dtb}+\frac{1}{4m}(\beta^n-\beta^{n+1}) \, . \nonumber \\ \label{eq:Eq_SV_5a}
\end{eqnarray}
Thus, Eqs.~(\ref{eq:Eq_SV_1a}), (\ref{eq:Eq_SV_2a}), and (\ref{eq:Eq_SV_5a}) constitute the identical SV form of the VE expressions.

\subsection{Leap-Frog G-JF}
This version of the Verlet method for Langevin equations comes with some flexibility in how
the method is expressed. We
take the starting point with the SV form given in Eq.~(\ref{eq:Eq_SV_1a}) and
introduce a reasonable definition of the half-step velocity
\begin{eqnarray}
v^{n+\frac{1}{2}} & = & \frac{r^{n+1}-r^{n}}{dt} \, . \label{eq:Eq_LF_1a}
\end{eqnarray}
We now use Eq.~(\ref{eq:Eq_LF_1a}) to replace $r^{n+1}$ in Eq.~(\ref{eq:Eq_SV_1a}) to obtain
\begin{eqnarray}
v^{n+\frac{1}{2}} & = & a\frac{r^n-r^{n-1}}{dt}+\frac{b\,dt}{m}f^n+\frac{b}{2m}(\beta^n+\beta^{n+1}) \, , \nonumber \\ \label{eq:Eq_LF_2a}
\end{eqnarray}
in which we can again apply Eq.~(\ref{eq:Eq_LF_1a}) and arrive at
\begin{eqnarray}
v^{n+\frac{1}{2}} & = & av^{n-\frac{1}{2}}+\frac{b\,dt}{m}f^n+\frac{b}{2m}(\beta^n+\beta^{n+1}) \, . \label{eq:Eq_LF_3a}
\end{eqnarray}
This equation is the half-time step velocity propagator, which is complemented by Eq.~(\ref{eq:Eq_LF_1a}) to yield a LF G-JF method
\begin{eqnarray}
r^{n+1} & = & r^n+dt \, v^{n+\frac{1}{2}} \, . \label{eq:Eq_LF_4a}
\end{eqnarray}
As was the case for the SV formulation of the method, this LF representation does not trivially incorporate
the natural initial conditions $(r^0,v^0)$. We therefore apply Eq.~(\ref{eq:Eq_LF_1a}) for $n=0$ in combination with 
Eq.~(\ref{eq:Eq_SV_2a}), resulting in
\begin{eqnarray}
v^{\frac{1}{2}} & = & bv^0+\frac{b\,dt}{2m}f^0+\frac{b}{2m}\beta^1 \, , \label{eq:Eq_LF_5a}
\end{eqnarray}
to be used, along with Eq.~(\ref{eq:Eq_SV_2a}), before applying Eqs.~(\ref{eq:Eq_LF_3a}) and (\ref{eq:Eq_LF_4a}) for $n>0$.
The proper integer-step velocity can be found from combining Eqs.~(\ref{eq:Eq_LF_3a}) and (\ref{eq:Eq_VE_1a}), resulting in
\begin{eqnarray}
v^{n+\frac{1}{2}} & = & av^{n-\frac{1}{2}}+2\frac{r^{n+1}-r^{n}}{dt}-2bv^n+\frac{b}{2m}(\beta^{n+1}-\beta^n) \, , \nonumber \\ \label{eq:Eq_LF_6a}
\end{eqnarray}
where we can then, again, use Eq.~(\ref{eq:Eq_LF_1a}) to obtain
\begin{eqnarray}
v^n & = & \frac{1}{2b}(v^{n+\frac{1}{2}}+av^{n-\frac{1}{2}})+\frac{1}{4m}(\beta^{n}-\beta^{n+1}) \, . \label{eq:Eq_LF_7a}
\end{eqnarray}
Thus, Eqs.~(\ref{eq:Eq_LF_3a}), (\ref{eq:Eq_LF_4a}), and (\ref{eq:Eq_LF_7a}) constitute a consistent LF method, where initial conditions $(r^0,v^0)$
are applied through Eqs.~(\ref{eq:Eq_SV_2a}) and (\ref{eq:Eq_LF_5a}).

Notice that while the given SV method is uniquely connected to the VE expressions in both the coordinate $\{r^n\}$ and its
evaluated velocity $\{v^n\}$, the development of LF is not unique, even if LF is
constructed to produce identical trajectories $(r^n,v^n)$ to those of VE and SV. The reason is the aforementioned somewhat ambiguous
choice in defining the half-step velocity $v^{n+\frac{1}{2}}$ in Eq.~(\ref{eq:Eq_LF_1a}), and the subsequent reconstruction of the
integer-time velocity $v^n$ in Eq.~(\ref{eq:Eq_LF_7a}). Thus, one can
exercise some freedom of choice in the velocity equations when using the LF expressions.

{\it For example},
a sensible alternative to the half-step velocity in Eq.~(\ref{eq:Eq_LF_1a}) may be
\begin{eqnarray}
u^{n+\frac{1}{2}} & = & \frac{r^{n+1}-r^n}{bdt} -\frac{1}{2m}\beta^{n+1} \label{eq:Eq_LF_rev_1} \, ,
\end{eqnarray}
where we use the symbol $u$ for the revised definition of the half-step velocity. Following the procedure starting from
Eq.~(\ref{eq:Eq_LF_1a}), we can develop the following alternate LF formulation, which also yields identical trajectories
$(r^n,v^n)$.
The two equations for half-step velocity and integer-step position become
\begin{eqnarray}
u^{n+\frac{1}{2}} & = & au^{n-\frac{1}{2}}+\frac{dt}{m}f^n+\frac{1}{2m}\beta^n \, . \label{eq:Eq_LF_rev_2} \\
r^{n+1} & = & r^n+b\,dt \, u^{n+\frac{1}{2}} + \frac{b\,dt}{2m}\beta^{n+1} \, , \label{eq:Eq_LF_rev_3}
\end{eqnarray}
with initiating half-step velocity $u^{\frac{1}{2}}$ and conversion to integer-step velocity $v^n$ written
\begin{eqnarray}
u^{\frac{1}{2}} & = & v^0+\frac{dt}{2m}f^0 \label{eq:Eq_LF_rev_4}\\
v^n & = & \frac{1}{2}(u^{n+\frac{1}{2}}+au^{n-\frac{1}{2}})+\frac{1}{4m}\beta^{n}\, , \label{eq:Eq_LF_rev_5}
\end{eqnarray}
respectively.
As mentioned above, all LF formulations of the thermostat yield identical trajectories for $(r^n,v^n)$ and therefore constitute
the same method regardless of the specific definition of the half-step velocity. Choosing which
one to use is entirely a matter of convenience. The defined half-step velocity is simply an auxiliary variable, which need not have any specific useful physical interpretation for the evaluation of thermodynamic quantities.

\section{Testing the method}
We exemplify the applicability of the G-JF method in the context of a simple biomolecular simulation
and we make comparison to results of the widely-used molecular dynamics codes
AMBER \cite{Cornell_1995}, LAMMPS \cite{LAMMPS-Manual,Plimpton_1995}, and GROMACS \cite{vdSpoel_2005}.
These packages employ different commonly used stochastic thermostats. More comprehensive discussions of
other thermostats can be found in recent references~\cite{G-JF,LM,Goga_2012}.
The purpose of the following
simulations is not to present new or ground-breaking results in biomolecular science; instead, we choose a simple and
well-understood representative model system that can illuminate, through simulations within well-established
MD packages, some key features of the new algorithm as implemented into one of the codes that is particularly
amenable to revisions.

\subsection{Simulation Details}
	 We performed classical molecular dynamics simulations of alanine
	 dipeptide (illustrated in Figure \ref{Eavg}a), a small and well-studied
	 biomolecule.
		 Intramolecular interactions among the solute atoms, including bond,
         angle, dihedral, and non-bonded energies, were modeled with an AMBER 
classical all-atom molecular mechanical
		 treatment \cite{Cornell_1995}, coupled with the recent
         \emph{ff12} parameter set \cite{Case_2012}, while the extra-molecular 
environment was treated as vacuum.

		 This model system was simulated at a target temperature of 300~K
         with four integration schemes for comparison:
		 {\it i}) The BBK thermostat \cite{Brunger_1984}, which is implemented and expressed within
		 AMBER 12 (\emph{ntt}=3)
         in the leap-frog Verlet formulation \cite{Loncharich_2004}; {\it ii}) the 
method of Schneider \& Stoll \cite{Schneider_1978}, as implemented in the LAMMPS simulation 
software \cite{LAMMPS-Manual} (with the  combination of the \emph{nve} and \emph{langevin} fixes); {\it iii}) a 
variation of the method of van Gunsteren \& Berendsen \cite{vGunsteren_1982,vGunsteren_1988}, implemented in 
GROMACS \cite{vdSpoel_2005} (as the \emph{sd} integrator \cite{Scott_1999}), which includes velocity rescaling;
and finally {\it iv}) the G-JF thermostat \cite{G-JF}.

The latter was
             implemented into AMBER through small modifications to
			 the AMBER 12 source code \cite{Case_2012}.  Since the
			 underlying LF Verlet integrator used in the BBK method is
			 consistent with Eqs.~(\ref{eq:Eq_LF_3a}), (\ref{eq:Eq_LF_4a}), and (\ref{eq:Eq_LF_7a}), revisions were only necessary
			 with regard to the fluctuation terms.  The memory framework was
			 modified to include space for the correlating noise term from the previous time step, corresponding
			 to $\beta^n$ in Eqs.~(\ref{eq:Eq_LF_3a}) and (\ref{eq:Eq_LF_7a}), to be cached for use in subsequent integration
			 steps.
         A Langevin damping coefficient of 10~ps$^{-1}$~was used throughout all 
the simulations.
         Simulation input data for LAMMPS and GROMACS runs were generated 
directly from AMBER-formatted coordinate and topology files using the free 
tools \emph{amber2lammps.py} (distributed with the LAMMPS source code), and 
\emph{amb2gmx.pl} \cite{amb2gmx}, respectively.

	 Data were collected from sets of ten independent 100~ps simulations
	 at several values of the integration time step, from 0.5--3.1~fs in increments
	 of 0.1~fs, and from 3.1--3.2~fs in increments of 0.01~fs.  Each simulation was
	 initiated from the same energy-minimized starting structure of the peptide.
	 For each independent simulation, initial atomic velocities were assigned
	 from a Maxwell-Boltzmann distribution appropriate to the target
	 temperature and the psuedo-random number chain was uniquely seeded.

	 First and second moments of the total potential energy distribution were
	 chosen as measures of the stability and accuracy with which
	 each integration method could reproduce the statistical-mechanical
	 properties of the physical system.  Average energy and its fluctuation
	 were computed based on samples at all integration steps.  The total
	 number of integration steps $N$ was varied based on the time step $dt$,
	 such that $N = \lfloor100~\mathrm{ps}/dt\rfloor$.

\subsection{Results} 
\begin{figure}
\includegraphics[width=8.5cm]{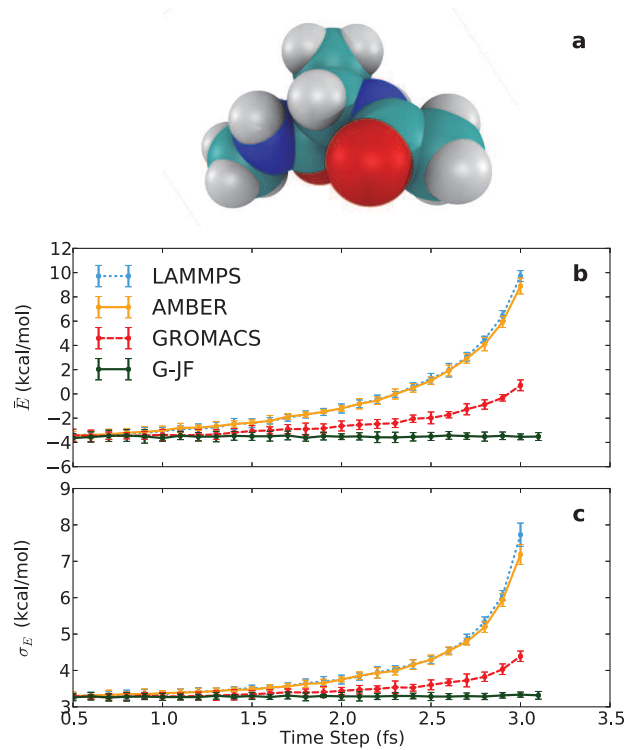}
 \caption{(a) Space-filling model of the simulated alanine dipeptide (Ac-Ala-NHMe)
     molecule. (b) Mean and (c) standard deviation of the potential energy in the 
model alanine dipeptide, computed with different molecular dynamics codes and 
various time steps.  Results are depicted as follows, top curve to bottom: 
Blue for LAMMPS; orange for AMBER; red for GROMACS; and green for AMBER with
the G-JF implementation.}
 \label{Eavg}
 \end{figure}

 Figures \ref{Eavg}b and c summarize the results and shows the average total potential energy and its
 standard deviation as a function of the simulated time step for all the 
integration methods.
 As expected, all methods give very similar results for both the average and 
standard deviation
 of the energies for as long as the time steps are smaller than about 1 fs.

However, for increasing time steps the unmodified contemporary codes start deviating from
the expected statistical values found for small $dt$. The two strongest deviations are found for the
BBK method, implemented in AMBER 12, and the thermostat of Schneider \& Stoll,
implemented into LAMMPS. The same deviating behavior, albeit seemingly with only one
third deviation, is found for the thermostat of van Gunsteren \& Berendsen, implemented
into GROMACS. In contrast, the G-JF method, implemented into AMBER 12 as described above,
shows the inherent feature of preserving the fluctuation-dissipation relationship for any time step.
In fact, for these simulations, the G-JF method can give statistical estimates at all time steps up to the
 stability limit that are indistinguishable from those at small time steps. The stability limit is here
 identified by adiabatically increasing the time step of a simulation until simulations produce
 anomalously high velocities, indicating that the Verlet integrator is no longer capable of
 producing a meaningful trajectory for the simulation. The existing stochastic thermostats seem
 stable up to 3.0~fs for this simulation, while the G-JF is stable for up to a slightly higher value 3.1 fs.

We close this section by arguing for the use of potential energy (and its fluctuations) as a measure for
the ability of the different thermostats to correctly sample the phase space corresponding to the target temperature.
A seemingly more straightforward test would be to consider the kinetic energy. However, as shown in
Ref.~\cite{G-JF} for the harmonic oscillator case, the computed velocity $v^n$, and thereby the kinetic energy,
is increasingly depressed for increasing time step {\em in any Verlet scheme}, regardless of the inclusion of a thermostat.
Thus, when evaluating a thermostat, measures of velocity and kinetic energy may not be appropriate
for determining the quality of statistical sampling. Further, this observation may hint at statistical sampling problems
arising from thermostats that involve ``velocity rescaling'' as a way to obtain a desired temperature.

\section{Discussion and Conclusion}
The newly developed G-JF stochastic Verlet thermostat has been applied to simulations of
a small, yet non-trivial and nonlinear, system representative of many
applications in molecular modeling. It has previously been analytically demonstrated that for linear
systems the new method yields exact statistical behavior of diffusion and
Boltzmann distributions for any time step leading to stable dynamics \cite{G-JF}. The simulations presented
here indicate that these attractive and robust statistical features are likely to remain in other complex and
nonlinear systems. Specifically, our present
simulations demonstrate that the statistical behavior of potential energy remains sound for
any time step up to the limit where the atomic
trajectories suddenly diverge. In contrast, available contemporary molecular dynamics
codes with other stochastic thermostats show
deviating behavior for increasing time step, indicating that the interpretation of thermodynamic
data from those algorithms must be done with caution and small time steps.
This seems to be the case for the three popular molecular dynamics codes that we have investigated here,
and is likely to be true also for other available MD simulation codes.
In order to facilitate comparison between the stochastic thermostats, we use the same AMBER force fields in all four sets of simulations. We emphasize that simulations with the G-JF thermostat have been completed by implementing
the new simple algorithm into an existing available code (AMBER 12), to ensure direct comparison
between the thermostats without any other differences in parameters or simulation details. Thus, based on the
preliminary tests and analyses, we suggest that
the algorithm presented here and in Ref.~\cite{G-JF} be implemented into existing
molecular dynamics codes for further use and evaluation. In order to facilitate such revisions, we have
explicitly provided the algorithm in all three commonly used formulations of the Verlet method.

\begin{acknowledgments}
The authors thank George~Batrouni, Daniel~Cox, Richard~Scalettar, and Rajiv~Singh for encouraging discussions.
This work was supported primarily by the US Department of Energy Project DE-NE0000536 000. Work was also supported by the Research Investments in the Sciences and Engineering 
(RISE) Program (UC~Davis) and US NSF Grant DMR-1207624.
\end{acknowledgments}


\begin{thebibliography}{99}
\bibitem{notice_on_methods} We notice that most recent methods are capable of reproducing correct
statistical properties for diffusion in a flat potential and approximate results for the harmonic oscillator. In contrast, the method given in \cite{LM} gives good statistical results for the harmonic oscillator, but does not  correctly reproduce spatial diffusion (as calculated from the Einstein relation) in a flat potential.
\end{thebibliography}
\end{document}